\documentclass[sigconf]{acmart}

\AtBeginDocument{%
  \providecommand\BibTeX{{%
    \normalfont B\kern-0.5em{\scshape i\kern-0.25em b}\kern-0.8em\TeX}}}

\copyrightyear{2021} 
\acmYear{2021} 
\setcopyright{rightsretained} 
\acmConference[CHI '21 Extended Abstracts]{CHI Conference on Human Factors in Computing Systems Extended Abstracts}{May 8--13, 2021}{Yokohama, Japan}
\acmBooktitle{CHI Conference on Human Factors in Computing Systems Extended Abstracts (CHI '21 Extended Abstracts), May 8--13, 2021, Yokohama, Japan}\acmDOI{10.1145/3411763.3451659}
\acmISBN{978-1-4503-8095-9/21/05}

\usepackage{array}
\usepackage{arydshln}
\begin{document}

\title{Ethical User Interfaces: Exploring the Effects of Dark Patterns on Facebook}


\author{Thomas Mildner}
\affiliation{%
  \institution{University of Bremen}
  \country{Germany}
}

\author{Gian-Luca Savino}
\affiliation{%
  \institution{University of Bremen}
  \country{Germany}
}

\renewcommand{\shortauthors}{Mildner and Savino}

\begin{abstract}
  Many researchers have been concerned with whether social media has a negative impact on the well-being of their audience. With the popularity of social networking sites (SNS) steadily increasing, psychological and social sciences have shown great interest in their effects and consequences on humans. In this work, we investigate Facebook using the tools of HCI to find connections between interface features and the concerns raised by these domains. Using an empirical design analysis, we identify interface interferences impacting users' online privacy. Through a subsequent survey ($n=116$), we find usage behaviour changes due to increased privacy concerns and report individual cases of addiction and mental health issues. These observations are the results of a rapidly changing SNS creating a gap of understanding between users' interactions with the platform and future consequences. We explore how HCI can help close this gap and work towards more ethical user interfaces in the future.
\end{abstract}



\maketitle

\section{Introduction \& Motivation}

During the rapid rise of social networking sites (SNS) and the number of people using them in a little more than a decade, an increase of mental health issues among their audiences has been recorded and published \cite{selfhout2009different, o2011impact, sagioglou2014facebook, van2008online}, while further studies looked at their impacts on social capital and connectedness \cite{allen2014social, ryan2017, ahn2013social}. Although participants of longitude studies self-reported a negative impact on physical and mental health when using Facebook \cite{shakya2017}, they still kept using the application regularly. A possible explanation can be derived from existing research on internet addiction ~\cite{kuss2011, Ryan2014, biolcati2018}. Christakis et al. has further linked an experienced decrease of future well-being in Facebook users to the disease in earlier works. ~\cite{christakis2009trapped, christakis2010internet}. Other researchers, however, did not find a reason to believe that SNS cause their audience to experience higher risks of mental health problems, such as depression \cite{Jelenchick2013FacebookDS}. Moreover, significant findings show that audiences of SNS may perceive higher levels of social connectedness \cite{ahn2013social, grieve2013face, sinclair2017facebook} and improved overall well-being with regards to reduced stress and social support \cite{nabi2013facebook}.

As people perceive and interact with SNS on the level of user interfaces, we believe that the perspective of the HCI community can add important insights to describe the relationship between a person and an interface with a focus on well-being. Recent research in HCI has already shown some interest in social media and their features. The `liking' behaviour of people on SNS, for example, has been investigated, noticing specific elements that influence the count of likes received \cite{jang2015}. With a focus on the social network Facebook, Wang et al. found that people often feel regret about certain content that they have shared which ultimately caused them disadvantages in social and professional areas \cite{wang2011}. In a follow-up study, Wang et al. implemented additional interface features that gave people the chance to review and correct content while displaying all recipients before publishing \cite{wang2013}. They found that most participants approved their features and thus demonstrated possibilities in which interface design can respond to people's concerns. A recent study from Andalibi et al. describes anxiety among interviewees concerning emotion recognition technologies that is expected to become a big part in SNS perhaps giving new possibilities to show fitting content \cite{andalibi2020}.

As a result of the fast changes and growth in technology, including SNS, society and research have fallen behind to provide ethical guidelines creating a `cultural lag'~\cite{gray2018, ogburn1957}. This lag is especially problematic when people are being misguided into doing something that they either did not expect or even tricked into actions with harmful results. With regards to Brignull's et al. work defining Dark Patterns \cite{brignull2015dark}, Grey et al. have showcased such occurrences in various digital interfaces. Often, Dark Patterns utilise knowledge about human psychology in combination with usable design to create deceiving design practices which do not have the user’s interests in mind~\cite{gray2018}. Because SNS want their audience to be recurring and promoting their service, it is unlikely that they want them to be associated with any form of harm. Thus, traditional Dark Patterns as defined by Brignull et al.~\cite{brignull2015dark} mostly seem unfit for SNS.

Prominent SNS, such as Facebook, Twitter, and Instagram, offer their services for free and generate their revenue through advertisement \cite{devries2012, kim2016}. The more time people spend on them, the more income can thus be generated. This motivates our research to take a look at design and interface strategies of SNS that may be linked to the decreased well-being through Dark Patterns or features similar to them. We, therefore, investigate Facebook, the currently largest social networking site, using common HCI methods like interface and user behaviour analysis. In this work, we present two studies: (1) In an empirical design analysis we identify and describe Facebook's key interface features and their changes in the past 16 years. (2) Through a survey with 116 participants, we analyse people's usage behaviour of Facebook with a particular focus on motivation, satisfaction, and privacy concerns. Based on these studies, we highlight scopes in which Facebook implements Dark Patterns, especially interface interferences. We further noticed a change in people's motivation to use Facebook due to privacy concerns and that those increase with age. Thus, especially younger generations might be exposed to harm through loose privacy settings and harmful features like `endless scrolling'. Eventually, we noticed a few general behavioural patterns but found worrying behaviour especially in individual cases, which we urge to further investigate in future research. With this paper we hope to spark the interest of the scientific community and contribute (1) an analysis of Dark Patterns in Facebook's desktop interface and (2) a discussion on how HCI can help to identify potential impacts of user interfaces to peoples' well-being based on the responses of 116 survey participants.

\section{Methodology}

Since their launch date in 2004, Facebook's interface received multiple updates. Meanwhile, additional features brought new options for people to engage with others and the application itself. In our research, we take a close look at Facebook's interface, its changes, and their effects on Facebook's audience while considering previous research in HCI, psychology, and social sciences. The presented study is split into two parts: (1) An empirical design analysis of Facebook's user interface and (2) a survey asking about people's Facebook usage.

\subsection{Empirical Design Study}
Two HCI researchers analysed Facebook's desktop user interface on an empirical level. To collect the necessary images of past UI iterations we used manual web crawling and collected screenshots of Facebook's desktop interface from a wide range of years and were crosschecking screenshots from different sources to ensure their accuracy. Further analysis focused mainly on Dark Patterns and privacy concerns. It included tracking certain UI-elements like the logout button and its positioning in the navigation bar as well as several iterations of the `account' menu. The understandings gained through this empirical approach led to the design of the Facebook user survey.

\subsection{Facebook Usage Survey}
To understand people's motivation, satisfaction, and privacy concerns when using Facebook, we created a questionnaire comprising of 30 questions including attention checks. The questionnaire consisted of `yes/no/maybe', 5-point Likert scale, and open-ended questions. We structured it in four sections: (1) demographics, (2) current/past motivation to use Facebook and most used features, (3) satisfaction of the time spent on Facebook and specific features, and (4) the participants' attitudes towards privacy concerns. We recruited participants through \textit{SurveySwap} and \textit{Reddit}, and hosted the surveys on Qualtrics~\cite{qualtrics2021}. We received 126 responses from which we excluded 10 due to failed attention checks or double entries. In the following results section, we analyse the data of these 116 survey respondents.

\section{Results} 

\subsection{Empirical Design Analysis}
Developed as a student network at Harvard University in 2004, Facebook received their most recent visual update in 2019. In the following paragraphs, we recorded notable changes to Facebook's desktop interface made within these years. For readability, the most prominent findings are visualised in Figure \ref{fig:interfaceRecords}. The following paragraphs present an overview of relevant changes to the logout button, the privacy settings, and the Privacy Checkup.

\subsubsection{Interface Interference}
Although being moved around the applications' interface, most of Facebook's features remained to be constant elements of the application. The logout button, for example, was part of the top-navigation until 2010. With an update, the button was moved into the `Account' menu, limiting discoverability. The privacy settings have experienced similar changes. Shortly after Facebook was opened to a wider audience, the privacy settings were moved from the navigation bar into the site's settings drop-down menu in 2008. In 2012, Facebook offered users an alternative on the first-level view, the Privacy Shortcuts. This feature allowed them to quickly access selected privacy settings and tools and was removed from the first-level interface with Facebook's latest redesign in 2019. 
Even though these changes could be natural consequences of responsive interface design, it is important to note that Facebook does benefit from users not logging out or sharing more information due to lighter privacy settings. This way Facebook is able to track their users across the web and use the information to create content and targeted advertisement. Intentionally moving these buttons to limit their discoverability and prevent certain user actions (i.e. logouts) would be considered interface interference in the Dark Pattern terminology~\cite{gray2018}. 

\subsubsection{Novel Dark Patterns}
During their latest redesign, Facebook introduced a new feature called Privacy Checkup. It allows users to edit privacy settings for pre-selected categories during a guided step-by-step process. Changes made in this feature directly affect the privacy settings, albeit with incomplete coverage. In addition to the privacy settings, users are further able to control ad-related settings. These also do not include the full spectrum of settings that are available in the general settings.
Especially ad-related settings can interfere with Facebook's advertisement strategies. By offering a guided settings feature Facebook is able to curate which settings users will manage. If this is used intentionally to keep users from certain settings, it could be seen as a novel way of interface interference. While placing all privacy settings behind several interface layers, Facebook actively offers a well designed but incomplete alternative to handle them.

\begin{figure*}[t!]
    \centering
    \includegraphics[width=0.70\textwidth]{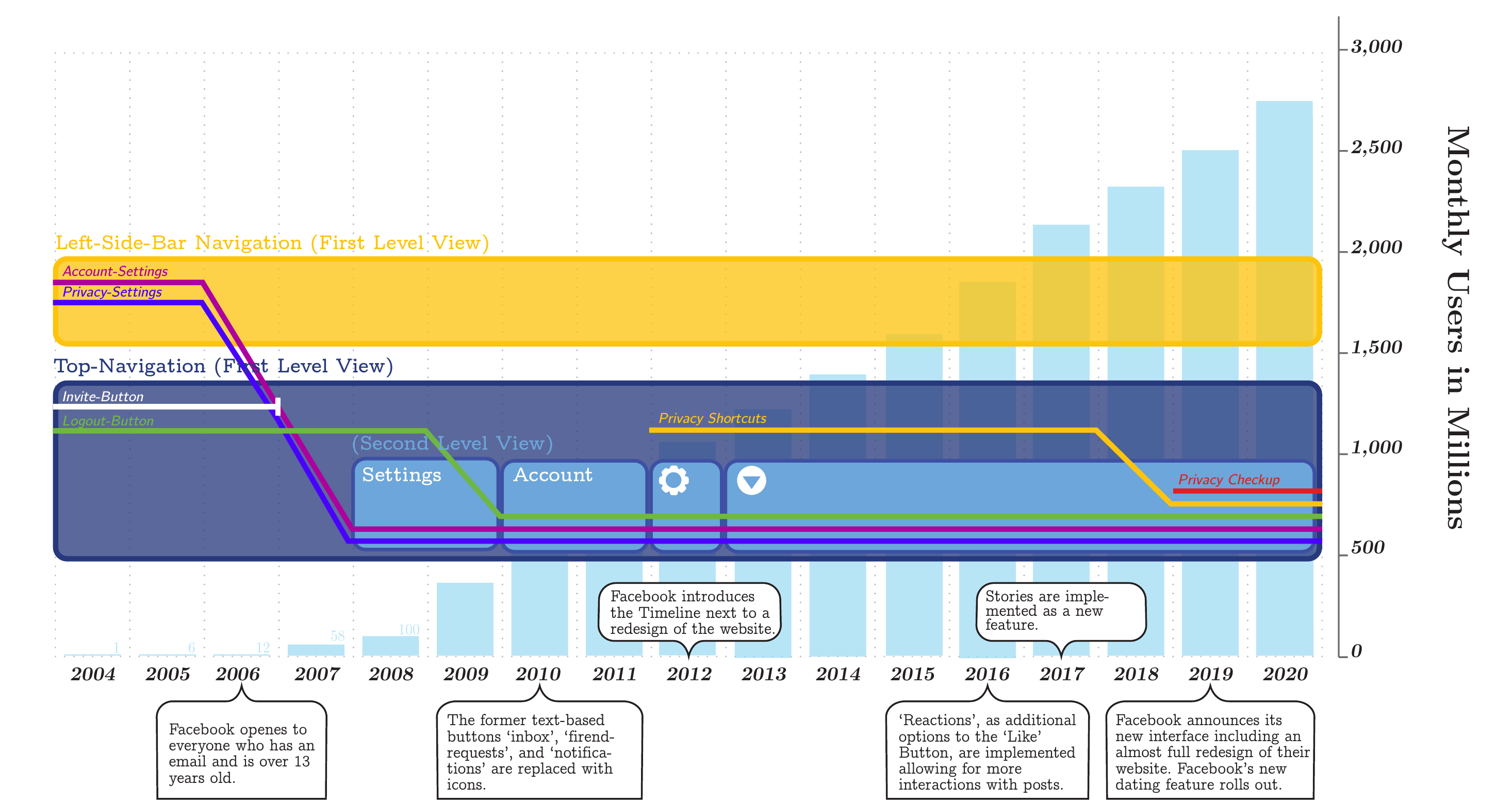}
    \Description[Figure 1 shows a diagram of Facebook’s interface changes between 2004 and 2020 with supplement data to highlight the increase of Facebook’s user count per year and descriptive texts for certain occasions.]
    {This diagram follows a time-axis from 2004 to 2020. Throughout the years, Facebook updated their user interface various times. The diagram describes interface elements as they were added to the application while following their position throughout multiple updates. Facebook’s first level views are mainly split into a side-bar and a top-navigation.
    In 2004, the side-bar included Account-Settings and Privacy-Settings. In 2008, these elements moved to the Top-Navigation into the second level-view of the settings menu. A descriptive text for 2006 reads: “Facebook opened to everyone who has email and is over 13 years old.” As a result of this change, an invite-button, located in the top-navigation since 2004, was soon removed in 2007. Also in the top-navigation, the logout button, which was also present since 2004, was moved into the second level view of the settings menu, now called ‘account’ menu.
    Another descriptive text for 2010 reads: “The former text-based buttons [for] ‘inbox’, ‘friend-requests’, and ‘notifications’ are replaced with icons”. 
    A descriptive text for 2012 states: “Facebook introduces the Timeline next to a redesign of the website” In this year, Facebook introduced another element – the Privacy Shortcuts. This feature was on the first-level view of the top-navigation until Facebook’s latest UI update in 2019. It was then moved into the ‘Account’ menu, now visualised by an icon of a downwards pointing arrow. This icon replaced a gear icon which was present in 2012. Also in 2019, Facebook introduced their Privacy Checkup next to the Privacy Shortcuts, a feature that allows users to make changes to pre-selected privacy settings.
    The diagram shows three  further texts: The first, a descriptive text for 2016 reads: “‘Reactions, as additional options to the ‘Like’ button, are implemented allowing for more interactions with posts.” The second text for 2017 reads: “Stories are implemented as a new Feature.” A last descriptive text for 2019 states: “Facebook announces its new interface including an almost full redesign of their website. Facebook’s new dating feature roll’s out.”

    For each year the respective user count is presented in a bar plot as supplement data. Facebook counted 1 Million users in their last quarter of 2004. In what looks like linear growth, the count increased to 2,740 Million users in the 3rd quarter of 2020. The user count for each year are as follows: 2004: 1; 2005: 6; 2006: 12; 2007: 58; 2008: 100; 2009: 360; 2010: 608; 2011: 845; 2012: 1056; 2013: 1228; 2014: 1393; 2015: 1591; 2016: 1860; 2017: 2129; 2018: 2320; 2019: 2498; 2020: 2740.
}
    \caption{This diagram visualises Facebook's interface changes between 2004-2020. With increasing popularity the interface was extended in multiple iterations. Because no official and complete data-set was present at the time of writing this work, the data of monthly Facebook users was collected from two sources. Firstly, the data for the years between 2004-2008 were retrieved from Facebook Newsroom \cite{facebook2021, sedghi2014}. As Facebook Newsroom no longer features their original data, an article from The Guardian that relies on the same source is cross-referenced for support. However, comparisons need to be tentative and taken with caution. Secondly, the data representing the years from 2008-2020 was gathered from Statista.com \cite{clement2020}.}
    \label{fig:interfaceRecords}
\end{figure*}

\subsubsection{Summary}
The goal of the empirical study was to find indications for Dark Patterns within Facebook's interface. Moving the logout button and privacy settings into drop-down menus can be classified as interface interference~\cite{gray2018}. Given that Facebook directly benefits from users not logging out (by being able to track them across the web as long as they are logged in), this can be a conscious choice to limit discoverability and thus prevent certain user actions (i.e. logouts). 
Traditionally, Dark Patterns cause some degree of harm to their audience. As SNS aim to keep theirs entertained and satisfied, this creates a contradiction to what is typically described by Dark Patterns. We did, however, find a novel way in which Facebook guides their users' actions in their interest. Past research has shown that managing privacy settings on Facebook is not an easy task for their users and most often does not result in the level of security that they expect~\cite{Liu2011}. Facebook now offers a feature which makes this process easier, but potentially still leaves the user with wrong expectations. This design choice leaves the impression of traditional Dark Patterns but is not yet covered by their terminology. 

\subsection{Facebook Survey}
The following results were extracted from the survey and present people's usage of Facebook, as well as our three main focal points on their motivation, satisfaction, and privacy concerns when using Facebook. Since our survey is exploratory, we primarily used descriptive statistics.

\subsubsection{Demographics}
Of the 116 respondents 46 ($40\%$) identify as male, 67 ($58\%$) as female, one identified as other, and two did not prefer to say. Their age ranged from 16 to 52 with a mean age of 26 ($SD = 7.4$). Respondents were situated in 29 different countries with roughly half coming from the UK ($n=19$), the US ($n=23$) and the Netherlands ($n=27$). Of all participants, 81\% ($n=94$) actually use Facebook currently and 19\% ($n=22$) do not. Of those who use Facebook, 74\% joined the network between 2005 and 2012. The majority (65\%) of respondents use Facebook on mobile more than on desktop, for 17\% it is the other way round. 18\% use it the same amount on both.

\subsubsection{Motivation}
To get an understanding of people's general motivation to use Facebook, we asked participants about their current and original motivation (why they initially signed up) as well as their most-used and most-regrettable features. The 94 active Facebook users were allowed to give multiple replies and named a total of 133 current motivations. We identified six main categories:  (1) 'Keeping in Touch with Family and Friends' (31\%); (2) 'Groups of Interest' (14\%); (3) 'News/Information' (13\%); (4) 'Messenger/Chatting Feature' (12\%); (5) 'Events' (8\%); and (6) 'Entertainment' (7\%). 21 replies (15\%) did not fit those categories and comprise of individual motivations. Regarding the original motivation (when first joining Facebook) respondents named a total of 88 which were sorted in the same six main categories:  (1) 'Keeping in Touch with Family and Friends' (41\%); (2) 'Groups of Interest' (3\%); (3) 'News/Information' (0\%); (4) 'Messenger/Chatting Feature' (6\%); (5) 'Events' (2\%); and (6) 'Entertainment' (3\%). Within the remaining 39 replies we identified two other large categories: (1) 'Like/Share' (20\%); (2) 'Peer Pressure' (11\%). 'Keeping in Touch with Family and Friends' remains an important feature of Facebook from back when people original started using it to today. While Facebook's 'Groups of Interest' feature is mentioned more often as a current motivation compared to the original one, 'Like/Share' seems to have declined being mentioned only once as a current motivation. The participants were further asked to state their most used and most regrettable Facebook features. The two most used features of a total of 181 features mentioned, were (1) 'Messenger' (29\%) and (2) 'Groups of Interest' (21\%). Other notable features comprised of: (3) 'Like/Share' (8\%); (4) 'Newsfeed' (8\%); and (5) 'Entertainment' (7\%). Additionally, participants mentioned 29 features they regret spending too much time on. People mostly regret spending time on 'Entertainment' (34\%) features on Facebook, followed by the 'News Feed' feature (24\%). 
Together 'Entertainment' and 'News Feed' makeup 15\% of people's most used features. Thus, they are among the lesser-used features compared to others. Still, some people use them while, at the same time, regret spending too much time on them.

\subsubsection{Satisfaction}
We further asked participants about their satisfaction with the time they spent on Facebook and whether they would like to change something about it. Most users were extremely (29\%) and somewhat (39\%) satisfied with the time they spent on Facebook. 20\% answered that they were neither satisfied nor dissatisfied and the remaining 12\% were somewhat dissatisfied. No participant was extremely dissatisfied which was mostly due to them wanting to spend less time on Facebook rather than more time. 49\% of respondents reported to sometimes find themselves longer on Facebook than they had planned to and 13\% about half the time or more often. Only 37\% never find themselves using it longer than they planned to. In contrast, 80\% of respondents never want to spend more time on Facebook than they already have. When asked directly, 14\% think they spend too much time on Facebook, 18\% think they maybe do, and 68\% do not think they spend too much time on Facebook.

The 32\% ($n=30$) who think that they do or maybe do spend too much time on Facebook listed the following reasons for why they think this is the case: 7 participants answered that they use Facebook to procrastinate and distract themselves from tasks they should be doing; 4 answered that it is addictive or has addictive features and that they would like to stop using it but still use it every day; 4 mentioned that they often mindlessly scroll and that the infinite scrolling behaviour of the timeline supports this habit. Other responses included that it is just entertaining, boring, or a waste of time. Finally, we asked all 94 participants if they would like to actually spend less time on Facebook. To this, 50\% answered that they do not want to spend less time, 21\% replied with maybe, and 29\% do would like to spend less time.

\subsubsection{Privacy Concerns}
In order to learn more about people's privacy concerns, our questionnaire comprised four directed question items. We distinguished between social interaction and business related incentives from Facebook by asking targeted questions in a 5-point Likert-Scheme. A majority of the participants generally want to be in control over the information that other people can see about them. For 50\% of them, this is an extremely important desire while 1\% do not find this control important at all. When asked about whether they actually feel in control over such information, 63\% agree while 23\% rather do not. When instead asking about the control over the information that Facebook uses for advertisement, 29\% find it extremely important whereas 6\% find it not important at all. We again asked about whether they feel in control over Facebook's use of their data for advertisement. While 13\% actually feel in control over the information that Facebook uses for advertisement, (definitely yes - 4\%), 70\% of our participants do not feel in control (definitely no - 34\%). This is further visualised in Figure \ref{fig:privacy-concerns}. Eventually, we asked participants to state their logout behaviour as well as their general concern about internet security. Most of the participants (79\%) never log out of Facebook, 7\% log out regularly, and 5\% do so always. When it comes to internet security, 73\% are generally concerned whereas 11\% are not.

\section{Discussion}

Research is divided on what effects social media and SNS have on people's mental health. While some results show small positive effects when spending time on social media, others yield small negative effects, or did not identify any effect at all. On one hand, various studies have found connections between increased depressive symptoms, that in occasions tragically resulted in suicides, and increased screen-time, new media, and SNS \cite{twenge2018}. On the other hand, researchers did not identify any such links \cite{coyne2020} or instead observed overall positive effects on people's well-being \cite{ahn2013social, grieve2013face, sinclair2017facebook}.

\begin{figure*}[t!]
    \centering
    \includegraphics[width=1\textwidth]{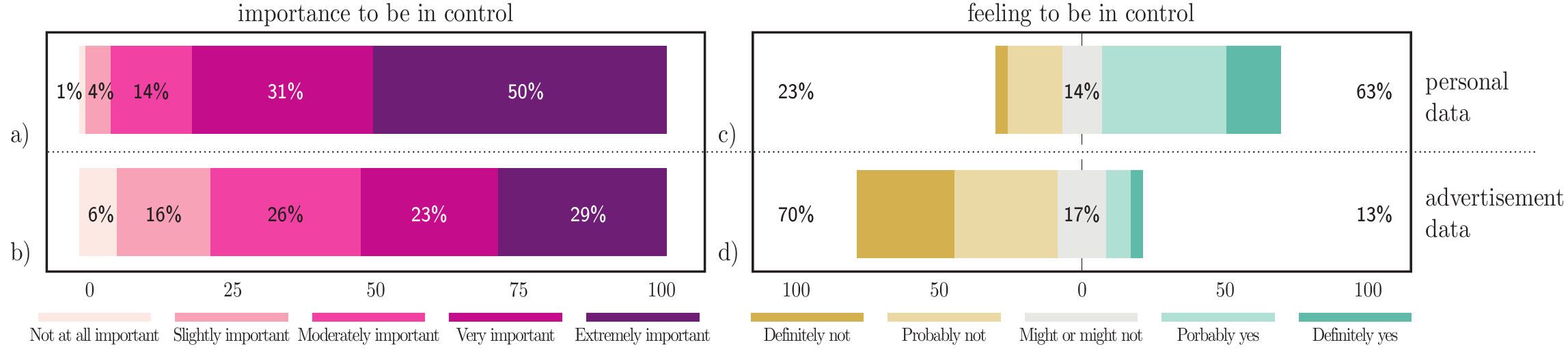} 
    \Description[Figure 2 compares four horizontal bar plots that show their results in a Likert-Scheme from 1-5. Questions asked about people’s privacy concerns when using Facebook while comparing their importance to control specific data and their feeling to be in control.]
    {This graphic shows four horizontal bar plots for the items: a) 'How important is it for you to be in control over the information other people can see about you on Facebook?'; b) 'How important is it for you to be in control over the information about you which Facebook uses for targeted advertising?'; c) 'Do you feel in control over the information other people can see about you on Facebook?'; d) 'Do you feel in control over the information about you which Facebook uses for targeted advertising?'
    The bar plots are arranged so a comparison between them can be done with respect to people's importance to be in control of their data as well as their feeling to be in control. It can be seen, that while 50\% of the participants find it extremely important to be in control over the information that others can see about them, only 29\% expressed the same importance regarding information Facebook uses for advertisement. In general, people found it less important to be in control over data used for advertisement than what others can see about them. 
    A similar shift can be observed for the items asking about peoples’ feelings to be in control. While 63\% of our participants feel to be in control over the information that others can see only 13\% feel in control over the information that Facebook users for advertisement. Meanwhile, 23\% say that they don’t feel in control over information that others can see while 70\% say that they don’t feel in control over the information used for advertisement.}
    \caption{This graphic shows the results of the questions: a) 'How important is it for you to be in control over the information other people can see about you on Facebook?'; b) 'How important is it for you to be in control over the information about you which Facebook uses for targeted advertising?'; c) 'Do you feel in control over the information other people can see about you on Facebook?'; d) 'Do you feel in control over the information about you which Facebook uses for targeted advertising?' of the survey.}
    \label{fig:privacy-concerns}
\end{figure*}

As so often in technology, SNS change quickly and thus require research to regularly re-evaluate their impact. HCI has the utilities to continuously analyse SNS and their effects on users with a particular focus on human-centred design. It can describe the relationship a person has with an interface and actively propose design guidelines to prevent possibly harmful design choices such as Dark Patterns.

In this work, we focused on this relationship between people and interfaces. We highlighted Facebook's continuous interface changes and noticed that design choices have led to potential Dark Patterns. In the presented empirical design study, we have identified two interface interferences. As can be seen in Figure \ref{fig:interfaceRecords}, Facebook has changed the location of the logout button as well as the link to the privacy settings multiple times while offering alternative but incomplete features. Although the cause for these changes may be reasoned through aesthetics, they result in unnecessary obstacles that users have to overcome. In the following discussion, we aim to create possible links between the presented studies. Our interface analysis only focuses on Facebook's desktop application while most participants of our survey preferred the mobile application. Thus, the conclusions drawn from these comparison should not be generalised as they reflect only on a smaller portion of our sample.

Even though they might not be intentionally developed, Dark Patterns can be used to evaluate the ethical principles of interfaces by assessing the difficulty of, for example, basic user actions. In the case of Facebook's logout button, it could explain why a majority of the participants never log out of Facebook, while also showing concern about internet security. Since Facebook uses cookies to track personal data within and outside their website \cite{facebookcookies2020}, our findings demonstrate an incongruity between people's privacy concerns and their actual activities when using SNS. As can be seen in Figure \ref{fig:privacy-concerns}, our results show that participants want control over the data that they share with others. This preference also exists for the data which Facebook uses for advertisement, although to a lesser extent. During their updates, Facebook chose to implement their Privacy Shortcuts and Privacy Checkup as alternatives to regular settings. With both, they provide their audience with controls over the general privacy settings. However, by limiting the number of options Facebook governs over the settings that their audience will adjust. Although such alternative interfaces are not covered by traditional Dark Patterns, they create interferences that allow Facebook to navigate their audience's decision-making without causing immediate harm.

Governing, but never harming, users makes it possible to keep their general satisfaction with Facebook high. Even though the majority of participants are generally satisfied with their time spent on Facebook, most of them spend more time than they actually plan to. Those who actually think they do spend too much time mention reasons like procrastination, distraction, mindless scrolling and even addiction. This shows a worrying development in which individual participants even realise their bad habits: "No time at all would be the ideal time, but the fact that I still open the app once a day shows a bad habit." P(102).  Facebook seems to keep general satisfaction up, but it is the individual person who develops problematic usage behaviours through features like endless scrolling (\emph{"infinite scroll makes me stay longer"} (P113)). Investigation of such features offer insights into how user interfaces may affect people's well-being.

An important part of this process is to understand people's incentives and motivations when using a service. By understanding what features people use or do not use or how their usage behaviour has changed over time, we are able to identify user groups and can connect them to behaviours and habits. Individual comments of the participants describe a perceptional change in their motivation to use Facebook. Sharing content, for example, made up 20\% of people's original motivation to create a Facebook account: \emph{"To [...] post about my social life"} (P44). Interestingly, it was only mentioned by one participant as their current motivation. A potential reason for this change could be the increased privacy concern as people grew up: \emph{"[I] used to be more active to share my life with others, but I don't anymore due to the privacy concern"} (P52). Our results show that with age participants' privacy concern also increased ($r = 0.23$, p-value = $0.029$). Unfortunately, we were not able to show statistically significant relationships between features or motivations and users' satisfaction. Still, this analysis helped us to understand what people do on Facebook and why they might quit the service or never even use it in the first place. We asked the 22 participants (19\% of our participants), who do not use Facebook actively, why they either stopped using the SNS or never even registered. Although their replies are highly individual, they foreshadow a link to the concerns of psychological and social science scholars, namely to well-being. P11 stated, that \emph{"it was bad for [their] mental health"} while P13 felt \emph{"judged by likes and comments"}. Although still a Facebook user, P24 wrote: \emph{"I felt bad about myself and my life"}, as a reason for why their motivation has changed. 

In this work, we explored how standard HCI methods like interface and user behaviour analysis can be used to identify potentially harmful aspects of Facebook's user interface and discuss how those could relate to effects on people's well-being. We do not find large numbers indicating general problems with SNS, but individual users show worrying usage behaviour and even specifically state mental health issues. As Beyens et al.~\cite{beyens2020effect} show, there can be vast individual differences, especially between adolescent users, in the way SNS affect their well-being. We, therefore, argue that researching the individual is just as important as the large scale studies which are already available. Through our approaches, we propose a first guideline for ethical interface design, namely considering users' motivation and expectations while testing user interfaces for Dark Patterns. Practitioners should further understand that high user satisfaction and usability is not necessary a result of ethical user interfaces.

\section{Limitations \& Future Work}
The empirical design analysis focuses only on Facebook's desktop interface, as it proved to be difficult to find enough authentic screenshots for all mobile versions due to various screen layouts based on different device sizes and operating systems (iOS and Android). An in-depth analysis of the mobile interface was thus out of scope for our work. Yet, the results of the survey show that most participants prefer Facebook's mobile application. While Facebook's latest interface update makes comparison between both interfaces easier, since both layouts show little to no differences, caution should be excersised when drawing conclusions. Future research should distinguish between all interface variations.

As with most studies utilising qualitative surveys, participants assessed their experience with Facebook. Comparing users' current usage behaviour with Facebook to their original motivation is necessary to understand how design changes may alter people's handling of a system. Nevertheless, the presented study asked participants to recall their experiences making these replies susceptible to recall bias. Moreover, chosen terminology, especially regarding one's well-being, are subjective. This subjective assessment allowed us to conduct in-depth analyses. We do, however, acknowledge that our sample was too small to find significant results in other parts of the evaluation. Especially for the results on participants' motivation larger sample sizes are necessary to fully understand their relevance. 

The presented research only focuses on Facebook since it is still the most widely used SNS. While more thorough research should continue to study Facebook, similar studies should investigate other SNS as the range of available SNS has grown in recent years. The research on the protection of human's well-being on SNS is an interdisciplinary effort. HCI adds a range of utilities that help to understand the interaction between people and SNS on the level of interface and technology. This community can thus support psychological and social sciences by actively impacting future design development and providing ethical guidelines while advancing its interdisciplinary efforts.

\section{Conclusion}

This paper presents two studies investigating the social medium Facebook. The first describes an empirical study analysing Facebook's continuous interface changes throughout recent years. We find cases of interface interference that Facebook has implemented with respect to the logout button and the privacy settings. In contrast to traditional Dark Patterns, these do not cause direct consequences but indirectly govern how they interact with the interface. Through a survey with 116 participants, the second study finds that users' motivation changed over time, making them share less personal data than they used to, due to increased privacy concerns. While being concerned about their private and ad-related data, users only partly feel in control about the amount they share. This feeling could be a result of the interface interference through which Facebook actively limits users' choices and guides their behaviour. This negatively impacts users (e.g. sharing information by not logging out), demonstrating an example for Dark Patterns on SNS. With this contribution, we explore different methods to show how HCI can contribute to investigating the well-being of social media users and hope to spark the interest of the scientific community to engage in this research.

\bibliographystyle{ACM-Reference-Format}
\bibliography{references.bib}

\end{document}